# Conformational switching of chiral colloidal rafts regulates raft-raft attractions and repulsions

Joia Miller[1], Chaitanya Joshi[1], Prerna Sharma[1,2], Aparna Baskaran[1], Gregory M. Grason[3], Michael F. Hagan[1], Zvonimir Dogic[1,4]

[1]*Department of Physics, Brandeis University, Waltham, MA 02454, USA*
[2]*Department of Physics, Indian Institute of Science, Bangalore 560012, India*
[3]*Department of Polymer Science and Engineering, University of Massachusetts, Amherst, MA 01003, USA*
[4]*Department of Physics, University of California, Santa Barbara, CA 93106, USA*

**Abstract:** Membrane-mediated particle interactions depend both on the properties of the particles themselves and the membrane environment in which they are suspended. Experiments have shown that chiral rod-like inclusions dissolved in a colloidal membrane of opposite handedness assemble into colloidal rafts, which are finite-sized reconfigurable droplets consisting of a large but precisely defined number of rods. Here, we systematically tune the chirality of the background membrane, and find that in the achiral limit colloidal rafts acquire fundamentally new structural properties and interactions. In particular, rafts can switch between two chiral states of opposite handedness, which dramatically alters the nature of the membrane-mediated raft-raft interactions. Rafts with the same chirality have long-ranged repulsions, while those with opposite chirality acquire attractions with a well-defined minimum. Both attractive and repulsive interactions are explained by a continuum model that accounts for the coupling between the membrane thickness and the local tilt of the constituent rods. These switchable interactions enable assembly of colloidal rafts into intricate higher-order architectures with unusual symmetries, including stable tetrameric clusters and "ionic crystallites" of counter-twisting domains organized on a binary square lattice. Furthermore, the properties of individual rafts, such as their sizes, are controlled by their complexation with other rafts. The emergence of these complex behaviors can be rationalized purely in terms of generic couplings between compositional and orientational order of fluids of rod-like elements. Thus, the uncovered principles might have relevance for conventional lipid bilayers, in which the assembly of higher-order structures is also mediated by complex membrane-mediated interactions.

**Significance Statement:** We describe hierarchical assemblages of colloidal rods that mimic some of the complexity and reconfigurability of biological structures. We show that chiral rod-like inclusions dissolved in an achiral colloidal membrane assemble into rafts, which are adaptable finite-sized liquid droplets that exhibit two distinct chiral states of opposite handedness. Interconverting between these two states switches the membrane-mediated raft interactions between long-ranged repulsions and attractions. Rafts with switchable interactions assemble into analogs of electrostatic complexation observed in charged particulate matter. These results demonstrate a robust pathway for self-assembly of reconfigurable colloidal superstructures, that does not depend on tuning the shape and interactions of the elemental units, but rather on the richness of the emergent and still poorly understood membrane-mediated interactions.

**Introduction**: When suspended in a bulk isotropic liquid, colloids interact by forces which have steric, electrostatic or entropic origins (1). In comparison, the same particles suspended in an anisotropic liquid crystalline environment, or at a fluid interface, distort their environment and thus acquire fundamentally different interactions. For example, colloids in a liquid crystal alter the nematic director field, generating topological defects, which lead to unique long-range interactions that drive the assembly of complex architectures (2-5). Colloids bound to oil-water interfaces provide another example in which long-ranged attractive or repulsive interactions are caused by the interface deformations. These in turn are determined by how the interface wets the colloids (6, 7). Lipid membranes provide a complex environment that combines both liquid crystalline-like anisotropy due to the alignment of the fatty chains and interface deformations due to effective two-dimensional confinement (8). Inclusions distort both the membrane thickness and the chain alignment, leading to interactions that are even more multifaceted than those observed for colloids in liquid crystals or on interfaces (9-12). However, the nanometer size of lipid membranes limits visualizing the nature of membrane-mediated interactions and associated assembly pathways (13-22).

Tunable depletion interactions enable the robust assembly of monodisperse rod-like molecules into colloidal monolayer membranes, structures which mimic many of the properties of the lipid bilayers yet are about two orders of magnitude larger (23-26). The one-micron thick colloidal membrane allows for direct visualization of inclusions within the membrane and the associated environmental deformations. For example, recent experiments have shown that chiral inclusions dissolved in a colloidal membrane of opposite chirality assemble into colloidal rafts, which are micron-sized deformable liquid droplets consisting of a large but precisely defined number of rods (27-29). Here, we show that colloidal rafts can assemble even in achiral membranes, and that such rafts can switch between two distinct chiral states. Furthermore, we show that interconverting between these two conformational states switches the effective raft-raft pair interaction from a long-ranged repulsion to a short-ranged attraction. These conformationally switchable self-assembled mesoscopic structures and interactions drive formation of diverse higher-order reconfigurable assemblages. Using theoretical modeling, we show that the complex interaction behavior can be explained by considering the coupling between the shape of raft interfaces and the tilt field of colloidal rods within the membrane. We find that a minimal symmetry-based model of phase separated and tilted membranes can describe most of the experimental observations. Our results demonstrate how properties of the membrane environment can be tuned to generate complex self-assemblages from structurally simple building blocks.

**Experimental Results**: We used M13KO7 and *fd* bacteriophage, which are both rod-like colloidal particles with ~7 nm diameter and 2.8 $\mu$m persistence length (30). M13KO7 is 1200 nm long, while *fd-wt* is 880 nm long. Addition of a non-adsorbing polymer to a virus solution induces attractive interactions, causing the viruses to assemble into colloidal membranes, which are liquid-like monolayers of rods, aligned lengthwise, with lateral dimensions reaching hundreds of microns (23, 24). Local twisting of rods as preferred by their intrinsic chirality is fundamentally incompatible with the global constraints of the two-dimensional membrane geometry. Consequently, all the rods within the membrane interior are forced into a higher-energy untwisted state, while rods within a twist penetration length of the membrane edge are free to twist and thus lower their energy (36). This frustration leads to chiral control of the membrane edge tension (35).

As evidenced from studies of bulk cholesteric liquid crystals, aligned viruses lower their interaction energy by slightly twisting with respect to each other (31, 32). A single amino acid mutation of the major coat protein turns the wild-type left-handed M13KO7 or *fd-wt* virus into a distinct filament class, denoted by M13KO7-Y21M and *fd*-Y21M, which are both stiffer than the wild-type rods and right-handed (33). Varying the ratio of Y21M-*wt* viruses controls the magnitude of the cholesteric pitch in bulk liquid crystals and the effective chirality of the colloidal membranes. In a weakly chiral limit the rods at the membrane's edge form alternating domains of left and right handed twist that are separate by point like defects (34). The rod chirality controls the effective line tension of the domains of either handedness which in most cases induces difference in the size of left- and right-handed domains (35). However, for achiral membranes the domains of either handedness have the same spacing. This criterion reveals that colloidal membranes composed of 63% Y21M and 37% *wt* rods are effectively achiral. Based on these findings we define the chiral parameter: $\phi_{ch} = \frac{r_{M13K07} - 0.37}{0.63}$, where $r_{M13K07} = \frac{n_{M13K07}}{n_{M13K07} + n_{M13K07\text{-}Y21M}}$, $r_{M13K07}$ is the fraction of left-handed rods, and $n_{M13K07}$ and $n_{M13K07\text{-}Y21M}$ are the concentrations of long rods. For achiral mixtures, $\phi_{ch} = 0$, while $\phi_{ch} = 1$ indicates maximally left-handed rod composition. Achiral colloidal membranes composed of fd-*wt* and fd-Y21M exhibit an edge instability that is consistent with the absence of rod twist (34).

Colloidal membranes comprised of a uniform mixture of rods of opposite chirality, such as M13KO7 and *fd*-Y21M, also force the constituents within the membrane interior to untwist. However, rods with opposite chirality and different lengths can form chiral colloidal rafts that allow the twist to penetrate the membrane interior and thus lower their free energy. These colloidal rafts are finite-sized equilibrium droplets of one-handedness and length that coexist with the background membrane of the opposite handedness and different thickness (27). The finite twist at a raft's edge decays into the membrane bulk, driving long-range repulsive interactions between the rafts (37).

To study the effect of chirality on the raft stability, we systematically lowered the net chirality of the background membrane by mixing two rods of equal length but with opposite chirality. We added short weakly-chiral right-handed *fd*-Y21M to the achiral long-rod background membrane. Surprisingly, despite the weak chirality, the short rods still robustly assembled into micron-sized rafts (**Fig. 1**). These rafts in isolation were structurally similar to previously studied rafts formed in the chiral limit, $\phi_{ch} = 1$ (27). However, there was one major difference. In the chiral limit, rafts experienced long-ranged repulsive interactions. Consequently, they exhibited liquid order at lower densities and colloidal crystals at higher densities. In contrast, rafts in achiral membranes formed clusters, while leaving other spaces void, thus suggesting attractive interactions (**Fig. 1D, Supplementary Movie 1**). Furthermore, at higher densities raft clusters formed square crystalline lattices, hinting at the presence of complex interactions (**Fig. 1E**).

We observed intriguing behaviors even at low densities, where only a few rafts interact with each other. For example, isotropic pairwise-additive attractions would yield trimer clusters with the shape of an equilateral triangle, in which the distance between any two units is determined by the position of the attractive minimum. Instead, three colloidal rafts formed obtuse and isosceles triangle-like structures (**Fig. 2A-C**). Such configurations remained stable over the entire sample lifetime and never transformed into an equilateral configuration (Supplementary Movie 2). Four-raft clusters assumed another unusual yet highly stable architecture that cannot be explained by simple attractions. Specifically, we observed a central raft that was surrounded by three outer rafts, arranged into an equilateral triangle (**Fig. 2D**, Supplementary Movie 3). These observations further suggest complex raft-raft interactions with an attractive component.

Motivated by these observations, we directly measured the pairwise potentials between colloidal rafts. Since short-rod rafts are repelled from a focused light, we used a time-shared optical trap arranged into a plow (27). Bringing two rafts into close proximity and shutting of the traps revealed two distinct behaviors: some pairs remained bound for the entire observation time, while other pairs drifted apart from each other over a matter of seconds (**Fig. 3A, B**). To quantify interactions we used the blinking optical trap (BOT) technique (38). We brought rafts together with the optical plows and quantified their subsequent trajectories. The laser was shuttered before taking any measurements to avoid the effects of trap-induced membrane distortions (**Fig. 3C**, Supplementary Movie 4). Acquiring many trajectories yielded two distinct transition probability matrices. One describes an exponentially repulsive interaction with a length scale of ~0.6 μm, which is similar to interactions in the previously studied chiral membranes (27) (**Fig. 3D**). The other corresponds to a ~6 $k_B$T attractive well with a well-defined minimum at 1.6 μm raft edge separation. The latter results were confirmed by tracking isolated attractive raft pairs, in equilibrium, over a period of time. These measurements generated the effective probability distribution function of the raft pair separation, $p(r)$. Inverting with the Boltzmann relationship: $\Delta V(r) = -k_B \text{T} \log p(r)$, yielded a similar potential, within an unknown constant due to an unidentified zero point (**Fig. 3D**). When using the BOT technique, a pair that was stably attractive in one experimental run could be repulsive in the next. Similarly, a previously repulsive pair could bind after the laser was shuttered. Rafts switched between two interaction types only when being manipulated with an optical trap. Left on their own they almost always remained in one state. These measurements demonstrate that rafts in an achiral membrane background exist in two conformational states with distinct attractive and repulsive interactions, and that they can switch between these two states.

To reveal the structural origin of switchable raft interactions, we visualized raft-induced distortions of the background achiral membrane using LC-PolScope, a technique that quantitatively images the sample optical retardance (39). For a monolayer membrane lying in the image plane, retardance is proportional to rod tilt away from the membrane normal (**Fig. 4A**) (36). Regions in which the rods are aligned along the membrane normal appear dark in LC-PolScope images. Rods tilted away from the normal have structural and optical anisotropy and thus appear bright (**Fig. 4A**). Previously, LC-PolScope microscopy visualized how twist penetrates the membrane interion over a characteristic lengthscale (36). When viewed with LC-PolScope, rafts in achiral membrane appeared bright, indicating local twist (**Fig. 4C, E**). However, there was no obvious difference in the appearance of the attractive and repulsive raft pairs.

Since it only measures the rod tilt away from the $z$-axis, LC-PolScope with normal incident illumination does not reveal the handedness of chiral rafts. To measure raft handedness, we instead illuminated the sample with the light incident at an angle $\theta$ with respect to the membrane normal (**Fig. 4.B**). Rods with positive local tilt in $y$-$z$ plane were then more aligned with the incident light. They thus had lower apparent retardance and appeared darker in the LC-PolScope image. In comparison, rods with negative tilt in the $y$-$z$ plane tilted away from the incident light at a larger angle. Having higher apparent retardance they appeared as brighter regions. Therefore, colloidal rafts imaged with a tilted angle LC-PolScope have an apparent asymmetry, which could be used to determine their handedness. Right-handed rafts appeared brighter on top and darker on the bottom, while the appearance of left-handed rafts was the reverse. Using combined tilted and normal incidence LC-PolScope revealed both the maximum raft twist and the twist handedness. All isolated rafts, as well as repulsive raft pairs, had internal right-handed twist, which is favored by the right-handed chirality of *fd*-Y21M (**Fig. 4C,D**). In contrast, each raft in an attractive raft pair had opposite handedness, despite one raft being in an higher-energy counter-twisted state (**Fig. 4E,F**). Normal-incidence LC-PolScope demonstrated that the maximum edge twist was of similar magnitude for rafts of either chirality (**Fig. 4C,E**). Trimers and tetramers also included a counter-twisted central raft, which was surrounded respectively by two or three right-handed rafts. The outer rafts repelled each other but were bound to the central raft with the opposite chirality (**Fig. 4G-J**, Supplementary Movie 5). The measured pairwise interactions can be used to predict the structure of raft trimers (**Fig. 2C**). However, these predictions are somewhat different from the measured structure, suggesting that raft interactions in dense clusters are not pairwise additive.

The colloidal membrane background is achiral and the rafts composed of *fd*-Y21M rods are weakly right-handed. Thus, right-handed twist should be thermodynamically favored by the intrinsic chirality of raft rods, while isolated counter-twisted left-handed rafts must be higher-energy metastable structures. Most isolated rafts were right-handed (**Fig. 5A**), and the number of counter-twisted rafts only became comparable to that of rafts with favorable twist at very high raft densities, such as in a square lattice. Using an optical trap, we isolated a counter-twisted raft by separating an attractive L-R pair, and subsequently observed its dynamics. The handedness, retardance and size of this raft remained unchanged throughout the entire observation period, demonstrating that the barrier to switching into lower energy twist state must be significant (**Fig. 5B-D, Supplementary Movie 6**). Using optical manipulation, we also assembled exotic clusters not found in equilibrium sample, such as a trimer with a lower-energy central raft that was bound to two outer counter-twisted rafts (Supplementary Movie 7).

To gain insight into the structural origin of switchable raft interactions, we have employed theoretical modeling. Previously developed continuum models for binary membranes (28, 29, 37) explained the assembly and repulsive interactions of chiral rafts in a membrane of opposite chirality. However, since the stability and direction of twist of rafts in these models is assumed *a priori* to be driven by chirality, they cannot explain the metastability of counter-twisted rafts observed here. Hence, the experimental observations of metastable rafts of both handedness imply that additional mechanisms must be at work to destabilize the rafts to spontaneous twist, even in the absence intrinsic chirality. The microscopic origin of this physical mechanism can be traced to the length asymmetry of the rods and the shape of a twisted raft-background domain edge, which drives the edge unstable to twist of either chirality, as will be described elsewhere. The key goal of the present study is to understand the physical mechanisms underlying the complex interactions between like and opposite twist domains.

To this end, we have extended the previously developed Ginzburg-Landau model to include a minimal model of edge-tilt coupling that can drive a raft unstable to spontaneous twist independent of the intrinsic chirality. The model accounts for the liquid crystalline elastic energy, coupling between variations in the local height of the membrane and the depletion interactions, and coupling between twist of the director field and compositional fluctuations. Membrane configurations are described by two fields: a director field $\hat{n}(r)$ that denotes the orientation of the rods relative to the membrane normal (assumed to be $\hat{z}$, with $\cos\theta = n_z$), and a concentration field parameter $\phi(r)$ which characterizes the local difference in the densities of short and long rods. The free energy is given by:

$$F/k_B T = \int d^2r \left[\frac{1}{2} K_1 (\nabla \cdot \hat{n})^2 + \frac{1}{2} K_2 (\nabla \times \hat{n})^2 - K_2 q(\phi)(\hat{n} \cdot \nabla \times \hat{n}) + Kaq_0\phi + \frac{1}{2}K_2 \sin^2\theta - \frac{\phi^2}{2} + \frac{\phi^4}{4} + \frac{\epsilon_\phi}{2}(\nabla\phi)^2 - \frac{\gamma}{2}\sin^2\theta (\nabla\phi)^2 + \frac{C_2}{4}\sin^4\theta\right] \quad (1)$$

The first three terms describe the Frank elastic energy associated with distortions of the director field, with $K_1$, $K_2$, and $K_3$ as the splay, twist, and bend elastic moduli. The local twist is coupled linearly to the concentration field $\phi$, $q = q_0 + a\phi$, where $q_0 \pm a$ corresponds to the preferred twist of the short-rods and long-rods, respectively. For the achiral background, $q_0$ is set to $a$. The fourth term describes the free energy cost of rod tilt arising from depletion interactions, with $C$ as the bulk depletion modulus which is proportional to the applied osmotic pressure. This term favors rods in the membrane interior to align with the membrane normal. The next three terms in even powers of $\phi$ account for the tendency of short and long rods to phase separate, with $\varepsilon_\phi$ being the line tension that penalizes the long-short rod interface. Everything up to this point was included in the previously developed model (29). The final two terms provide a mechanism for the raft edge-twist instability. While these terms should be expected due to the most generic coupling between tilt and composition, here we note that such terms can arise due to the exclusion of the depletant from a surface layer around the membrane, which leads to a free energy penalty $F_A = \Pi aA$, where $\Pi$ is the osmotic pressure of the depletant, and the excluded layer has a volume $aA$, with $a$ the depletant radius and $A$ the membrane surface area. The surface area increases due to variations in the membrane height according to $A = 2\int d^2r \sqrt{1 + (\nabla h)^2}$. The membrane height depends on both the local tilt and composition, $h(r) = t_\phi n_z$, with $t_\phi$ the composition-averaged half length of the rod - $t_\phi$ interpolates linearly between $t_s$ and $t_l$ as $\phi$ goes from $+1$ to $-1$. Incorporating the dependence of membrane height on local composition and expanding to second order in $\phi$, results in a term $-\gamma \sin^2\theta (\nabla\phi)^2$, with $\gamma \sim \Pi$. Notably, this term favors spontaneous twist and is independent of chirality. It thus serves a proxy for an edge-tilt coupling that renders rafts unstable to twist. The final $\sin^4\theta$ term must be added for stability, with $C_2$ as an adjustable parameter.

We tested the ability of our model to describe the experimental observations. We set $K_2=K_3=C=1$ and varied the splay elastic constant, $K_1$, and the surface tension, $\gamma$. Geometrical arguments suggest that $K_1 > K_2$ and $K_2 = K_3$ (40). First, we performed quasi-static calculations to calculate raft stability. In these calculations we fixed the radius and composition profile of a raft, and calculated the equilibrium director field by minimizing the free energy (**Eq. 1**). To determine the dependence of the free energy on raft size, we performed this calculation over a range of raft radius values. We found a range of $K_1$, $\gamma$ values over which right-twisted rafts with finite radius are stable, and counter-twisted rafts are metastable. Above a threshold value of chirality the metastable counter-twisted state disappears, consistent with previous studies with chiral background membranes which do not exhibit counter twisted domains.

Next, we performed an analogous procedure to calculate the interactions between raft pairs. We imposed a composition profile corresponding to two rafts with fixed radius and separated by distance $d$, and then optimized the free energy with respect to the director field. Our model predicts repulsive and attractive interactions for like-twisted and opposite-twisted raft pairs, respectively (**Fig. 6C,D**). Comparison of the theoretical results with experiments shows that the model captures most key features: stable twisted rafts independent of chirality, metastability of counter-twisted rafts at low chirality, and attractive/repulsive interactions between pairs of opposite/like twisted rafts. Importantly, we could not find a parameter range in which the model simultaneously captures all of these features without including an edge-tilt coupling, supporting our hypothesis that such an effect is essential for the switchable interactions observed in the experiments. We identify one discrepancy with experiment – the theory does not predict the short-range repulsions of oppositely-twisted rafts. This limitation may arise either from an additional cost of through-thickness density variation in splayed regions that form between opposite twist domains or additional thermodynamics of compositional fluctuations, both of which are neglected in the present model will be investigated in a future work.

To understand the origin of the switchable interactions, we first consider the repulsive interactions between rafts of the same chirality in an achiral membrane background (**Fig. 6C**). We consider the twist field within and surrounding two right-handed interacting rafts. Due to the coupling of the local membrane thickness to the local rod twist, each raft distorts the membrane-polymer interface in its immediate vicinity. This is reminiscent of interactions between interfacially adsorbed colloids, which generally attract each other since bringing particles closer together reduces the overall area of the distorted interface (6). Based on such reasoning one might expect that colloidal rafts will also experience attractive interactions, driven by the effective interfacial distortion energy. However, our model predicts, and we also experimentally measure, strong repulsive interactions. This can be explained by the intrinsic coupling between the membrane height and the rod twist, which is absent in simple liquid-liquid interfaces. We define the local tilt field, $\theta(x,y)$, with the origin of the coordinate system at the midpoint between two rafts separated by distance $d$. The twist at the edges of a pair of right-handed rafts, designated an R-R pair, is $\theta(-d/2,0) = -\theta_0$ and $\theta(d/2,0) = \theta_0$, were $\theta_0$ is the maximum twist angle at the raft boundary (**Fig. 6A**). Consequently, the rods in the background membrane between the two rafts have to distort from $-\theta_0$ to $\theta_0$ over a distance $d$, and the twist field crosses zero at the midpoint, $\theta(0,0) = 0$. In essence, rods at the midpoint between rafts must untwist, regardless of the raft separation (red line in **Fig. 6E**). Pushing two rafts closer together forces a sharp untwisting and height variation. This increases the interfacial-twist cost, creating large gradients of twist in the inter-raft region, and thus increasing the strength of the repulsive interaction. In contrast to interfacial colloids, bringing rafts together cannot reduce the total area of the deformed interface, and thus does not reduce repulsions. In the previously studied chiral limit ($\phi_{ch} = 1$), the inherent chirality of the left-handed membrane background and the right-handed rafts generated repulsions because bringing rafts together reduced the preferred twist of the inter-raft region (29). In contrast, for an achiral membrane background ($\phi_{ch} = 0$), the chiral contribution to the interaction energy is small (**Fig. 6C**). In this limit the increase in surface area and the associated excluded volume becomes the primary contribution to the repulsive interaction energy (**Fig. S3**).

The continuum model also explains the attractive interactions between the L-R raft pairs. Unlike a pair of right-handed rafts, the tilt field at the inner edges of a left- and right-handed (L-R) raft pair is in the same direction, that is $\theta(-d/2,0) = \theta(d/2,0) = \theta_0$ (blue line in **Fig. 6E**). This removes the geometric constraint that requires rods in the inter-raft region to untwist completely at the midpoint, leading both to less total twist deformation and a smaller surface area (**Fig. 6F**). When compared to an R-R raft pair, the elastic energy of the L-R pair increases due to the counter-twisted state of one of the rafts. However, this energy increase is more than compensated by the decrease in the twist deformation that is associated with removing the constraint of rod untwisting in the inter-raft region (**Fig. 6D**). As the separation of an L-R pair increases, the rods surrounding the rafts twist back to the membrane normal and any gains in excluded volume disappear (**Fig. 6G**). This generates an attractive well when rafts are close enough to share their respective twist fields. LC-PolScope images reveal that the inter-raft region of a bound L-R pair is significantly more twisted than a twist field of an isolated raft (**Fig. 7**).

To experimentally explore the limits of stability of the attractive L-R pairs, we systematically increased chirality of the membrane background, $\phi_{ch}$. This should decrease the stability of the counter-twisted rafts, since in this case both rods within the raft and outside have to twist against their preferred handedness. We studied the stability of L-R pairs in chiral colloidal membranes ($\phi_{ch} = 0.4$). Using an optical tweezer, we switched one raft to a counter-twisted state and formed a L-R raft pair. Such pairs remained stably bound indicating attractive interactions, but the counter-twisted raft shrunk over time until the pair fell apart (**Fig. 8A-B**, Supplementary Movie 8). At this point the two rafts diffused away from each other. Soon thereafter the originally left-handed raft started to increase in size again, implying that it had switched to the energetically more favorable right-handed conformation (**Fig. 8C**). This indicates that the range of attractive interactions between oppositely twisted rafts is independent of the background membrane chirality, while stability of the counter-twisted rafts does depend on the chirality of the membrane background. Surprisingly, we found that counter-twisted rafts are stabilized by multiple L-R bonds, even at $\phi_{ch} = 0.4$. Specifically, we observed stable tetramers under these conditions and the higher-energy central raft remained stable indefinitely (**Fig. 9A-D**). With increasing background chirality, the average size of the central raft shrank, while the outer rafts grew, to minimize the energetically unfavorable left-handed twist (**Fig. 9E**). Despite different raft sizes, the equilibrium separation of attractive pairs does not vary with changing $\phi_{ch}$, providing further evidence that the attractive interactions do not depend on the membrane chirality (**Fig. 9F**).

Rafts form a variety of structures depending on short rod density and dextran concentration. The ratio of right-handed to left-handed rafts depends on raft density. The inherent preference for right-handed internal twist is only overcome by the cost of the background membrane deformation in the presence of at least one other raft. As such, the few rafts that form in membranes with a low short-rod density are unlikely to be left-handed. At high raft densities, the membrane deformation is minimized by an equal number of left- and right-handed rafts assembled into a square lattice.

Between these two extremes, rafts assemble into heterogeneous structures of various sizes determined by the number of left-handed rafts (Supplementary Movie 1). Alternatively, rafts can assemble into chain-like structures in which links of alternating left- and right-handed rafts are joined by highly twisted necks (**Fig. S1A,B**). This is structurally different from the chains formed by rafts in a highly chiral background, in which each raft link has the same right-handed twist, and the rod twist is instead minimized at the necks that join the rafts (**Fig. S1D**). At intermediate background chirality, the link size is anisotropic and large right-handed rafts alternate with smaller left-handed rafts (**Fig. S1C**). As the dextran concentration increases, rafts become unstable in favor of bulk phase separation between long and short rods. This transition to bulk phase separation is preceded by a narrow phase space in which the left-handed rods in the achiral background mixture wet the rafts, possibly due to their chirality or comparatively low stiffness (**Fig. S2**, Supplementary Movie 9).

**Discussion**: Our work demonstrates an intricate multi-step assembly pathway of rod-like particles in the presence of depletant interactions. In a first step, isotropic bidisperse rods phase separate from the background polymer solution and form monolayer colloidal membranes. In the second step, the unique properties of the nascent two-dimensional membrane drive lateral phase separation of short and long rods, and assembly of colloidal rafts, which are liquid-like monodisperse clusters of inclusions that contain a large but well-defined number of short rods. Specifically, the micron-sized rafts we study are comprised of about ~20,000 rods, and can exist in two distinct chiral conformations. The higher-energy counter-twisted state is stabilized, at least in part, by attractive bonds formed with rafts of opposite chirality. In the final step, complex membrane-mediated inter-raft interactions, which are determined by the raft internal conformations, drive assembly into higher-order hierarchical structures such as highly stable tetramers and crystallites with square lattices.

Our work unifies diverse concepts from soft matter physics. By adding polymers to rod-like liquid crystal forming viruses, we induce assembly of colloidal monolayer membranes, which have important similarities and some distinctions with conventional lipid bilayers. Introducing an immiscible component into such membranes leads to the formation of finite-sized colloidal rafts, which have the appearance of colloid-like particles but differ in a fundamental aspect. Colloidal rafts are liquid-like deformable structures that maintain their finite-size despite continuous exchange of constituent rods with the background membrane. Therefore they are adaptable, self-healing assemblages that can sense their local environment, adjusting their structure and interactions accordingly. In this aspect they resemble thermodynamically stable surfactant micelles. Increasing the raft concentration leads to complex clusters, whose formation is driven by emergent membrane-mediated interactions. These clusters are significantly more complex than those formed by spherical colloids under simple depletion interactions (41).

Observations of membrane-mediated assembly of colloidal rafts are intriguing from multiple perspectives. First, they demonstrate that molecular chirality provides a robust platform for rational engineering of geometrically frustrated assemblages, wherein interactions between chiral elemental units favor local packing motifs that are incompatible with uniform global order, thus generating finite-sized structures, that unlike micellar assemblies are vastly larger than sub-unit dimensions (e.g. rod diameters) (42, 43). Second, the interactions that we have elucidated depend only on generic properties of the membrane, such as coupling between the membrane interfacial area, the membrane thickness, and the local tilt of the constituent rods (or aligned hydrophobic chains in a lipid bilayer). Therefore, interactions similar to those studied here could play a role in conventional lipid bilayers, and thus be relevant for the assembly of biological membrane inclusions. Third, there is an increasing emphasis on assembling complex architectures that go beyond traditional hard-sphere crystals. The typical approach towards this goal involves developing new synthesis methods for assembly of "patchy" building blocks with complex shapes and interactions. Our work demonstrates a different route toward this goal, which uses building blocks with simple shapes. When suspended in locally structured environments, these building blocks can acquire complex interactions that drive their assembly.

Extensive work has focused on elucidating the quantitative relationship between microscopic interactions and macroscopic phase behavior, and this has guided the rational development of patchy particles. Our experiments demonstrate the richness and diversity of interactions that emerge in membrane-like environments. It thus provides impetus to use theoretical and simulation tools to fully explore the manifold of all possible interactions that can emerge in membrane-like environments. Once developed, such a theoretical framework would fully elucidate both the potential and the limitations of membrane-mediated self-assembly.

**Materials and Methods:**

**Biological Purification and Sample Prep** For our experiments, we mixed filamentous bacteriophage of various contour lengths, persistence lengths, and chirality. These mixtures were composed of *fd-wt* and M13KO7 bacteriophage which are 880 nm and 1200 nm long respectively. Both rods have a diameter of 6 nm, 2.8 μm persistence length and form a cholesteric phase with a left-handed pitch. We also used the Y21M mutants of both strains which are similar in size to the wild type versions, but have a 9.9 μm persistence length and form a right-handed cholesteric (33). We purified all strains using standard biological procedures (44). The purified virus solution generally contains end-to-end dimers and multimers that favor smectic stacks rather than monolayer membranes. We removed multimers through isotropic-nematic phase separation (23). All viruses were suspended in a 20 mM Tris-HCl buffer (pH 8.0) to which 100 mM NaCl had been added to screen electrostatic repulsion between rods.

For fluorescence imaging, we labeled the primary amines on M13KO7 with DyLight 488, and the primary amines on *fd*-Y21M with DyLight 550 (45). There are approximately 3600 possible labeling sites on M13KO7 and 2700 labeling sites on *fd*-Y21M, of which we labeled less than 2% to ensure that the fluorescent markers would not affect system behavior. In all samples, we mixed the longer M13KO7 and M13KO7-Y21M at predetermined ratio. Previous studies demonstrate that a mixture of 37% M13KO7 and 63% M13KO7-Y21M exhibits no effective chirality (34). We added *fd*-Y21M to this mixture at number ratios between 10% and 50% to form short-rod rafts within membranes.

After adding a non-adsorbing polymer (dextran molecular weight 500 kDa, Sigma Aldrich) to the virus mixture solution at 40 mg/ml concentration, we injected the solution into a chamber formed by a coverslip and glass slide with a Parafilm spacer. Both the coverslip and slide were cleaned with a hot soap solution (1% Hellmanex) and coated with an acrylamide brush to prevent membrane adhesion on the glass coverslip surface (35). We sealed the chamber with optical glue (Norland). The total virus mixture concentration varied between 0.5-1.0 mg/ml.

**Microscopy** We used an inverted microscope equipped for differential interference contrast (DIC), fluorescence, LC-PolScope, and phase contrast imaging. All images were taken with a 100x oil immersion objective (Plan Fluor NA 1.3 for DIC and fluorescence, Plan Apo NA 1.4 for phase) and recorded on a cooled CCD camera (Andor Clara, Neo, or iXon).

Measurements of rod tilt within membranes were taken using LC-PolScope (39). LC-PolScope uses quantitative polarization measurements to find the local birefringence of a sample. The birefringence is displayed as an image in which the pixel intensity is proportional to birefringence, which itself is proportional to the local tilt of the rods due to the liquid crystalline nature of the membrane phase. When the membrane lies flat on the sample substrate, the rods in the membrane bulk are generally aligned along the axis of the incoming light, and the birefringence is at a minimum, leading to a low signal.

For blinking optical trap measurements, optical trap configurations were generated by time sharing a laser beam (4W, 1064 nm, Compass 1064, Coherent) using a pair of orthogonally oriented paratellurite acousto-optic deflectors (Intra-Action). The laser beam was projected onto the back focal plane of an oil-immersion objective (1.4 numerical aperture, 100X PlanApo) and focused onto the imaging plane. The multiple trap locations were specified by custom LABVIEW software. Because *fd*-Y21M-enriched rafts are shorter than the membrane background, the traps function as a plow which we use to push two rafts together or drag them apart. This enables us to watch the evolution of a raft pair after it is initialized away from its equilibrium separation. Raft separations were measured as a function of time, once the traps had been switched off, using standard video tracking methods (46). We created attractive raft pairs by pushing two rafts close together and then causing splay in the membrane by shifting the focus of the traps upwards in z. When the membrane relaxed, one raft of the pair twisted into the meta-stable left-handed state. The rafts remained in the bound state for the sample lifetime. The time lapse between successive frames was 500 ms and the exposure time was 50 ms.

The Blinking Optical Trap measurement is based on the fact that probability of the raft separation being $r_j$ at time $t_j = t_i + \Delta t$ can be determined by $p(r_j) = \sum_i P(r_i, r_j, \Delta t) p(r_i)$ where $P(r_i, r_j, \Delta t)$ is the transition probability for a raft pair separated by $r_i$ at time $t$. We find $P$ experimentally by binning pair trajectories by the initial and final separations for each time point. The equilibrium probability vector $\rho(r)$ is a steady state solution to the above equation and thus can be calculated as an eigenvector of the transition probability matrix (38). Alternatively, raft pairs were tracked in equilibrium conditions over long times to calculate an effective $\rho(r)$. In both cases, we used Boltzmann statistics to calculate the effective interaction energy from the calculated equilibrium probability: $\Delta F = -k_B \text{T} \log(\rho(r))$.

**Acknowledgements:** We acknowledge support of NSF-MRSEC-1420382 (to M.F.H., A.B. and Z.D.), NSF-DMR-1609742 (to Z.D) and NSF-DMR-1608862 (to G.G.). We also acknowledge use of the Brandeis MRSEC optical microscopy and biosynthesis facility supported by NSF-MRSEC-1420382.

**Figures:**

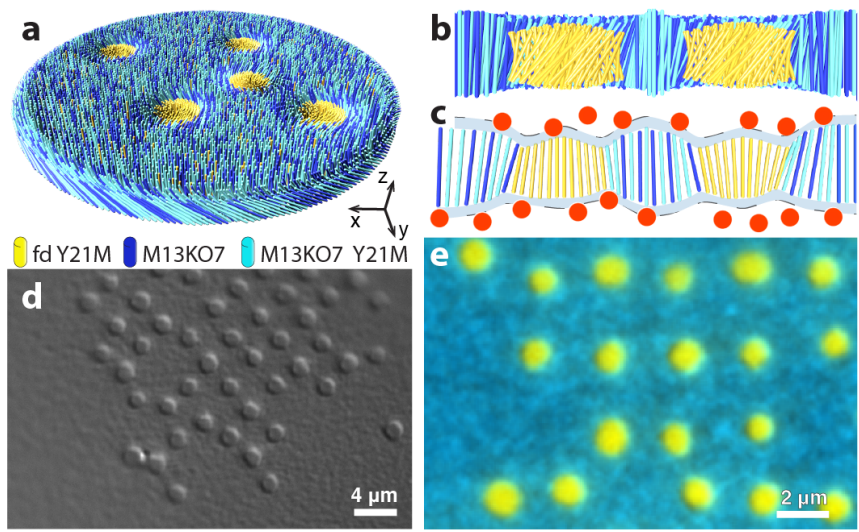

**Figure 1: Self-assembly of rafts in an achiral membrane.** (a) Schematic illustration of a colloidal membrane composed of an achiral mixture of long left-handed rods (M13KO7, dark blue), long right-handed rods (M13KO7-Y21M, light blue), as well as a small fraction of short right-handed rods (*fd*-Y21M, yellow); short rods form micron-sized rafts suspended in the membrane background. (b) x-z cross-section of a colloidal membrane containing a raft pair. (c) The rods in a colloidal membrane are held together by the osmotic pressure exerted by the enveloping polymer suspension indicated in red. The presence of a colloidal raft leads to variations in the membrane thickness. The depleting polymers are excluded from the membrane interior and from a layer adjacent to the membrane surface (indicated in grey). The latter leads to an effective surface tension that disfavors height variations. (d) DIC micrograph of a dense raft cluster coexisting with an empty background membrane suggests the presence of attractive raft-raft interactions. Raft clusters tend to form square lattices. (e) Fluorescence image of a cluster of rafts with square-like ordering observed in an achiral membrane. Raft-forming short right-handed rods (*fd*-Y21M) are indicated in yellow. The background membrane is composed of an achiral mixture of long left- and right-handed rods (respectively, M13K07 and M13K07-Y21M). Left-handed long rods are labelled and shown in blue.

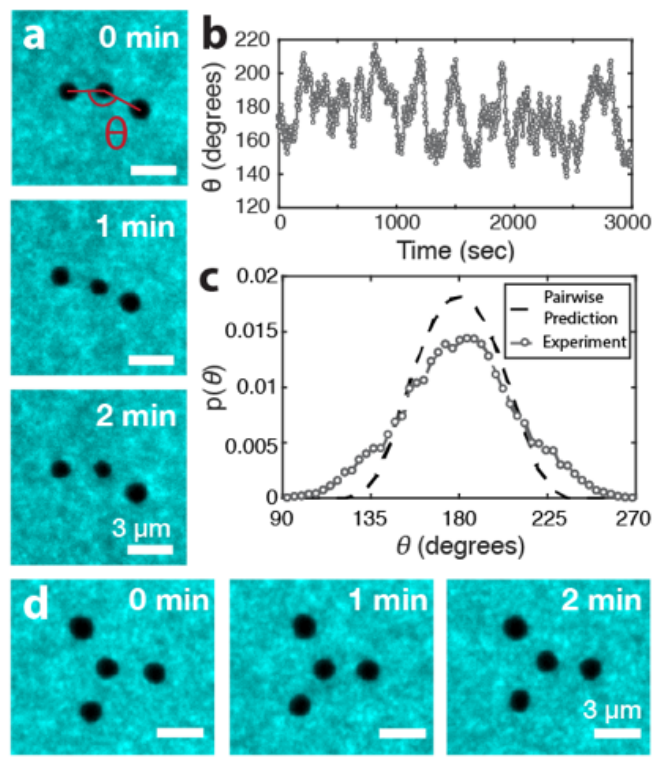

**Figure 2: Rafts form stable trimers and tetramers. (a)** A three-raft cluster forms a stable trimer that assumes the shape of an isosceles triangle, with the large angle denoted as $\theta$. **(b)** Plot of angle $\theta$ extracted from a fluctuating trimer. **(c)** Probability distribution function $p(\theta)$, extracted from a time-series of a fluctuating trimer. **(d)** A four-raft cluster forms a distinct and highly stable tetrameric structure, wherein three outer rafts are evenly spaced around a center raft.

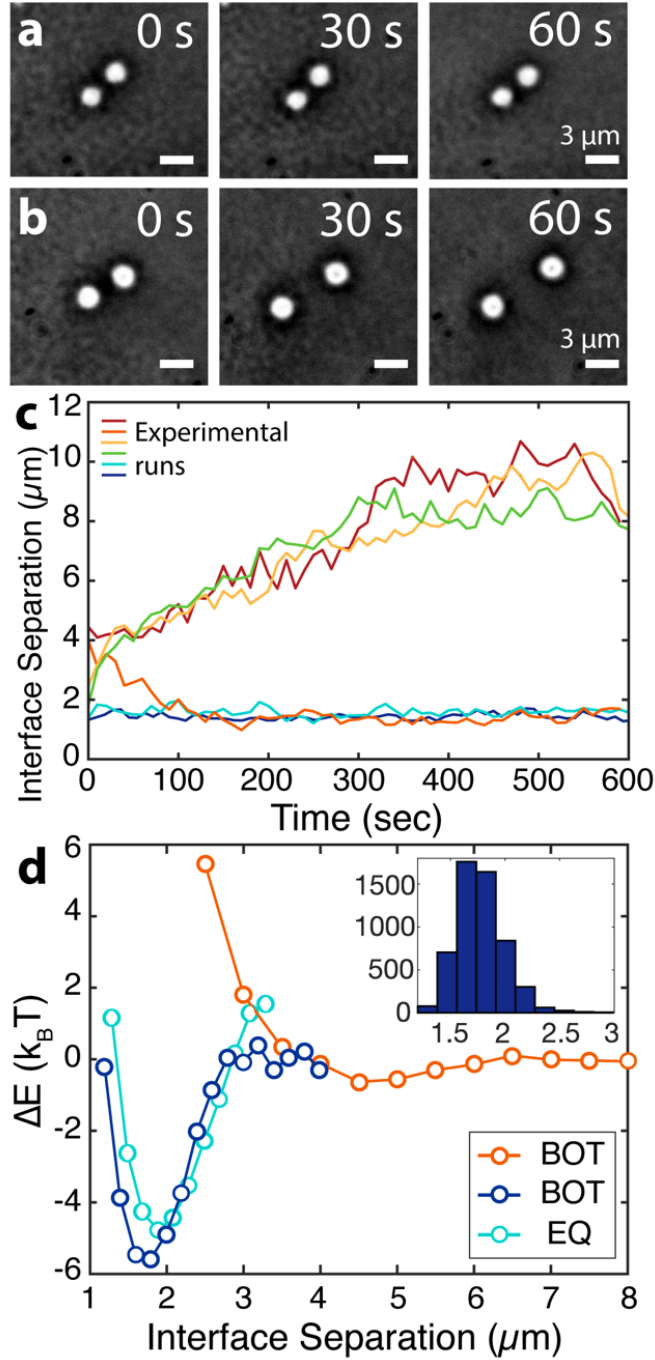

**Figure 3 Raft interactions can switch between attractive and repulsive. (a)** Upon being brought into close proximity with an optical plow, two rafts form a stable pair. **(b)** For identical conditions, a different raft pair repels each other. **(c)** Trajectories of several raft pairs brought into close proximity with optical plows. The data illustrates the existence of two categories of raft interactions that are either attractive or repulsive. **(d)** Effective interaction potentials for attractive

and repulsive rafts that are extracted from the raft trajectories using the blinking optical trap technique (BOT). Stable pairs have a potential minimum at 1.6 µm edge separation; unstable pairs have an exponentially repulsive potential, characterized by a ~0.6 µm lengthscale. Inset: The histogram of raft separations for an attractive raft pair. Attractive interactions extracted from equilibrium fluctuations agree with the BOT measurements to within an arbitrary constant.

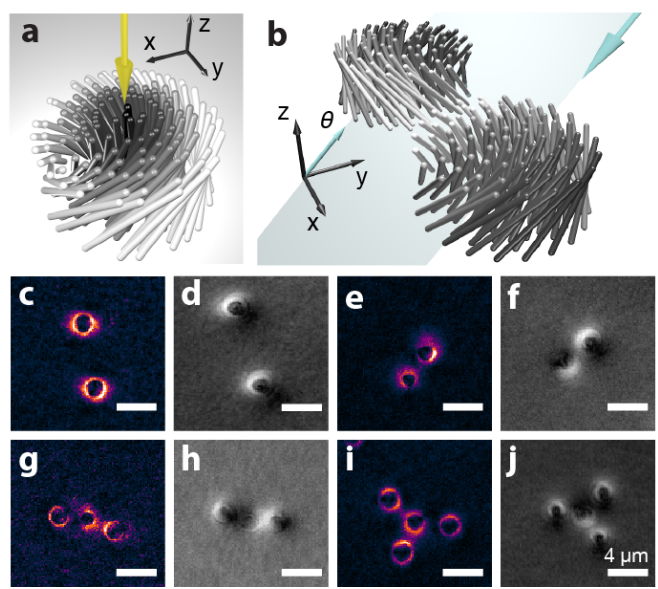

**Figure 4. Raft twist determines their interaction. (a)** Schematic of a twisted raft viewed with LC-PolScope. At the raft center rods are aligned with the incoming light, leading to zero retardance signal. Increasing the tilt away from the raft center leads to brighter signal in LC-PolScope. Normal illumination does not distinguish between the handedness of raft twist. **(b)** Tilting the incoming light along the *x*-axis with angle $\theta$ causes rods tilted along or against the incoming light to have different effective retardance values. When viewed with LC-PolScope, rods titled with angle $-\theta$ are bright while rods titled with angle $\theta$ appear dark. For right-handed rafts, this corresponds to bright signal at the back of the raft and dark at the front. These intensities are opposite for left-handed rafts, thus allowing us to differentiate between handedness of the raft twist. **(c, e)** Normal incidence LC-PolScope images of an attractive and repulsive raft pair. **(d, f)** Tilted incidence LC-PolScope shows that both rafts are bright at the top in a pair of repulsive rafts, while rafts are bright at the top and bottom in an attractive pair. Hence, repulsive and attractive pairs have respectively the same and opposite handedness. **(g)** Normal incidence LC-PolScope of a stable trimer. **(h)** Tilted LC-PolScope reveals that the central raft has opposite handedness from the two outer rafts. **(i)** Normal incidence LC-PolScope of a stable tetramer. **(j)** Tilted LC-PolScope reveals that the central raft has opposite handedness from the three outer rafts.

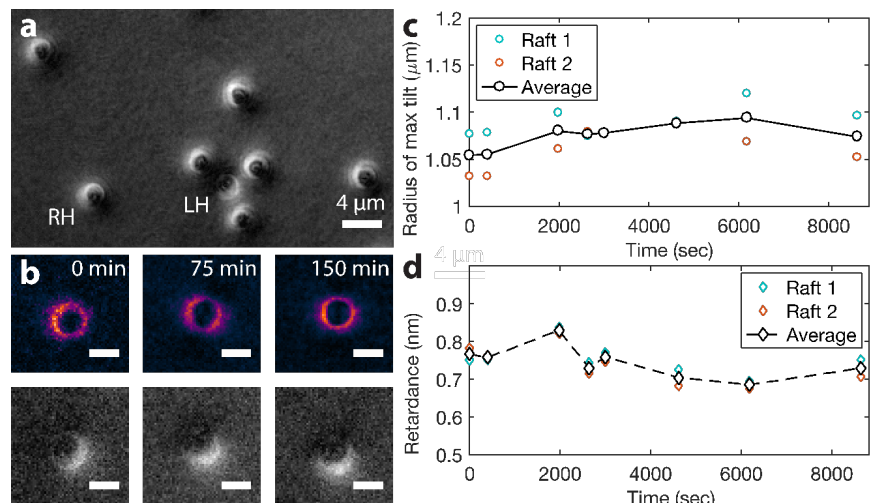

**Figure 5: Isolated counter-twisted rafts are metastable in achiral membranes. (a)** Titled incidence LC-PolScope of isolated rafts and a tetramer illustrates that isolated rafts have favorable right-handed twist. **(b)** Time sequence of an isolated counter-twisted raft shown in both normal and titled incidence LC-PolScope. **(c)** The size of a counter-twisted raft does not change with time. **(d)** The maximum tilt as measured by retardance of a counter-twisted raft does not change with time.

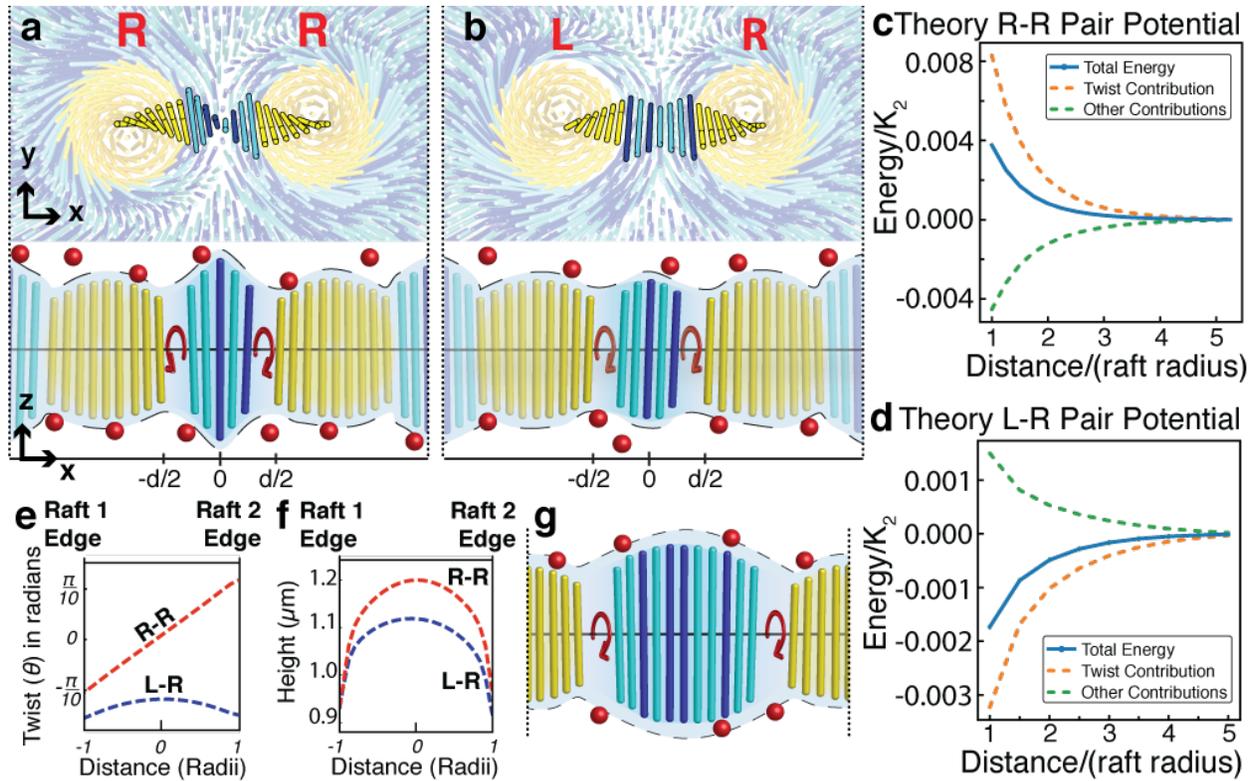

**Figure 6. The theoretical model shows that raft attractions arise from minimizing surface area in left- and right-handed pairs. (a-b)** Schematic representations of rod twist between rafts from above and from the side for pairs of rafts with the same (R-R) and opposite (L-R) twist. **(b)** edge twist direction. Model results for energy cost vs. distance between rafts in terms of their radii show surface costs dominate in both cases, leading to repulsive interactions for Right-Right pairs **(c)** and attractive interactions for Left-Right pairs **(d)**. **(e)** Schematic representation of twist relaxation for non-interacting rafts. **(f)** and **(g)** show the theoretical twist and height profiles respectively between two rafts given a separation distance equal to 2.0 raft radii. Parameters used in the calculations are: $K_1 = 3.0, K_2 = K_3 = C = 1$, and $\gamma = 0.8$.

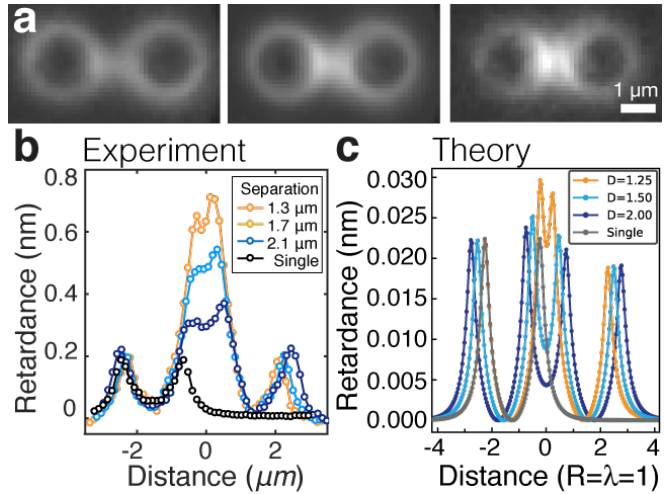

**Figure 7. Twist increases between L-R raft pairs. (a)** LC-PolScope images reveal the twist field of L-R raft pairs. The retardance field for separations of 2.1 μm, 1.7 μm, and 1.3 μm shows increasing retardance between rafts as the edge separation decreases. **(b)** Plot of the retardance along the midplane for each separation. **(c)** Model predictions of twist between attractive rafts also show increasing retardance as raft separation decreases. Distances are measured in units of the raft radius, which is equal to λ, the twist penetration length for short rods.

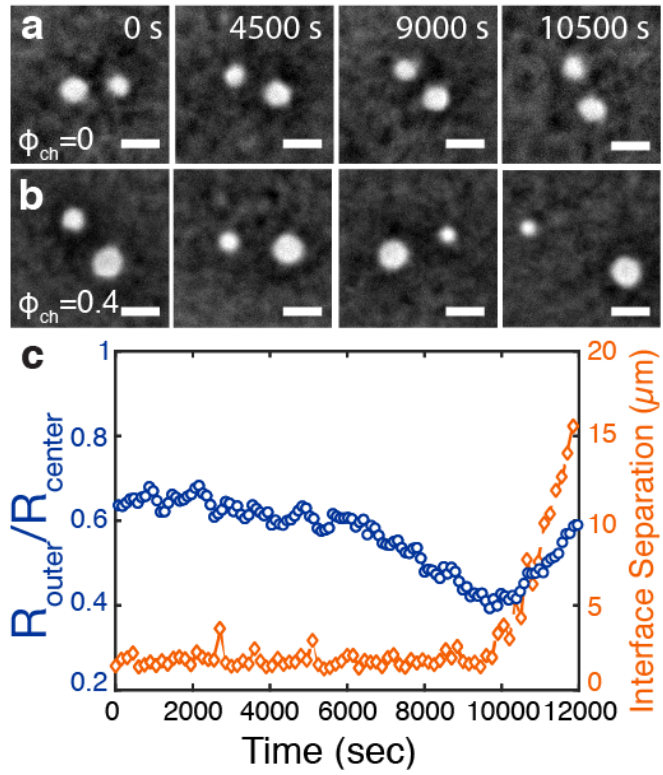

**Figure 8: A weakly chiral membrane background leads to metastable raft pairs. (a)** In an achiral membrane background, attractive raft pairs are stable over the sample lifetime. **(b)** Increasing the left-handed chirality of the background membrane ($\phi_{ch} = 0.4$) causes the counter-twisted raft to slowly shrink over time until the pair unbinds. Once unbound, the counter-twisted raft switches handedness and recovers its original size (Supplementary Video 8) **(c)** Time evolution of the ratio of the raft radii (blue circles) and edge separation (orange diamonds). Scale bars are 2 μm.

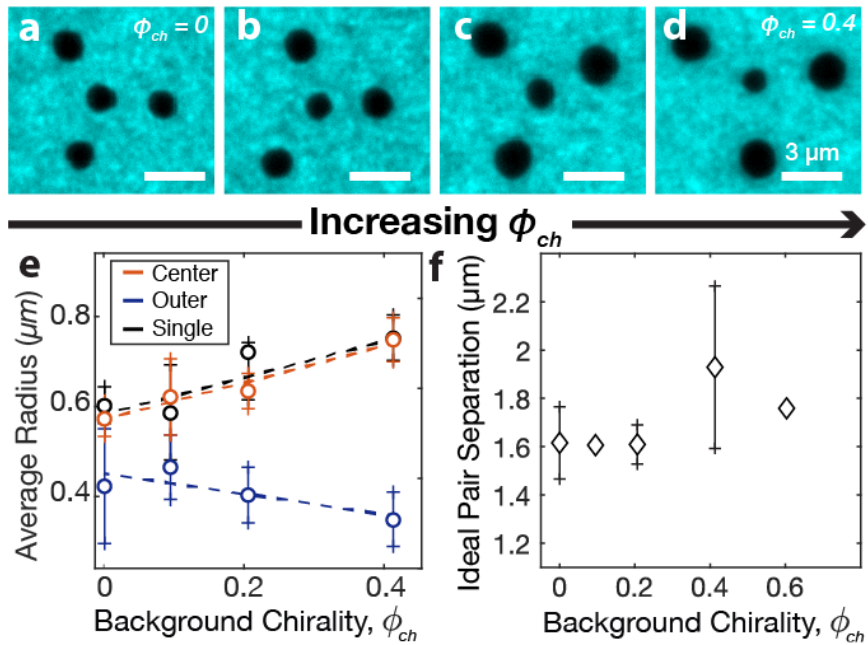

**Figure 9. Multiple L-R bonds stabilize counter-twisted rafts. (a-d)** Structure of stable raft tetramers formed in colloidal membranes at different background chirality values. **(e)** With increasing background chirality ($\phi_{ch}$), the size of the counter-twisted central raft shrinks and the outer rafts grow. The raft size of outer rafts in tetramer clusters is the same as those of isolated rafts. **(f)** The preferred length of a L-R bond does not depend on the background chirality.

**Supplementary Information Text**
**Theoretical Methods**

**Equilibrium Director Field Computations:** The equilibrium configuration for the free energy (Eq.(1) in main text) can be found by solving for $\frac{\delta F}{\delta \hat{n}} = 0$ and $\frac{\delta F}{\delta \phi} = 0$. We set $K_1 \neq K_2 = K_3$, and we use the units in which the unit length is $\sqrt{K_2/C} = \lambda = 1$. With these conditions, the free energy and its derivative become:

$$\beta F = \int d^2 r \left[\frac{1}{2} K_1 (\nabla \cdot \hat{n})^2 + \frac{1}{2} K_2 (\nabla \times \hat{n})^2 - K_2 q(\phi)(\hat{n} \cdot \nabla \times \hat{n}) + Kaq_0\phi + \frac{1}{2} K_2 \sin^2 \theta - \frac{\phi^2}{2} + \frac{\phi^4}{4} + \frac{\epsilon_\phi}{2}(\nabla \phi)^2 - \frac{\gamma}{2} \sin^2 \theta (\nabla \phi)^2 + \frac{C_2}{4} \sin^4 \theta\right] \quad \text{[S1]}$$

From here it follows:

$$\frac{\delta \beta F}{\delta \hat{n}} = -K_1 \nabla (\nabla \cdot \hat{n}) + K_2 \nabla \times \nabla \times \hat{n} - 2K_2 q(\phi)\nabla \times \hat{n} + K_2 \hat{n} \times \nabla q(\phi) + [K_2 - \gamma(\nabla\phi)^2 + C_2 \sin^2 \theta](n_x \hat{x} + n_y \hat{y}). \quad \text{[S2]}$$

We assume the concentration fields for the rafts to be circularly symmetric, with radial profile $\phi(r) = \tanh \frac{r-R}{\sqrt{2\epsilon}}$, with an interface width of $\sqrt{2\epsilon}$ (1). Here we have assumed that the corrections to the surface tension arising from the edge-tilt coupling term do not significantly change the interface width since the angles are small. The equilibrium configurations can be found by solving $\frac{\delta F}{\delta \hat{n}} = 0$ for $\hat{n}$.

For the calculations involving a single raft in the background membrane, the $\phi$ profiles are radially symmetric, and calculations show that the equilibrium $\hat{n}$ profiles are radially symmetric and (to a good approximation) splay-free. This allows us to simplify the calculation of $\hat{n}$: we define $\hat{n} = \cos\theta(r) \hat{z} + \sin\theta(r) \hat{\varphi}$ and solve just for the scalar field $\theta(r)$ instead of the vector field $\hat{n}(r)$. In this case, Eq.1 becomes:

$$\beta F = \int 2\pi r \, dr \left[\frac{1}{2} K_2 \left(\left(\frac{d\theta}{dr}\right)^2 + \frac{\sin 2\theta}{r} \frac{d\theta}{dr} + \frac{\sin^2 \theta}{r^2}\right) - K_2 q(\phi)\left(\frac{d\theta}{dr} + \frac{\sin 2\theta}{2r}\right) + Kaq_0\phi + \frac{1}{2} K_2 \sin^2 \theta - \frac{\phi^2}{2} + \frac{\phi^4}{4} + \frac{\epsilon_\phi}{2}(\nabla\phi)^2 - \frac{\gamma}{2} \sin^2 \theta (\nabla\phi)^2 + \frac{C_2}{4} \sin^4 \theta\right] \quad \text{[S3]}$$

$$\frac{\delta \beta F}{\delta \theta} = -2\pi K_2 \frac{d}{dr}\left(r\frac{d\theta}{dr}\right) + \pi K_2 \sin 2\theta \left(r + \frac{1}{r} - r\gamma(\nabla\phi)^2\right) + 2\pi K_2 \left(q + r\frac{dq}{dr}\right) - 2\pi K_2 q \cos 2\theta + 2\pi r C_2 \sin^2 \theta \sin 2\theta. \quad \text{[S4]}$$

We solve $\frac{\delta \beta F}{\delta \theta(r)} = 0$ using the finite element boundary value solver FEniCS (2, 3). For all results shown in this article, we set the short rod twist pitch $q_s \lambda = 0.02$, long rod twist $q_l \lambda = 0$, and the lengths of the rods as $t_s = 0.88 \, \mu m$ and $t_l = 1.20 \, \mu m$. For the plots of interaction energy as a function of distance (Fig. 6, main text), we used $K_1 = 3.0$ and $\gamma = 0.8$.

**Testing stability of counter-twisted rafts**: For positive $q_s > 0$ and $q_l = 0$, a right-handed raft has $\theta_+(r) > 0$ in the equilibrium solution. For there to be a stable counter-twisted raft, there should exist a second solution for $\frac{\delta \beta F}{\delta \theta(r)} = 0$ with $\theta_-(r) < 0$. We employ two methods to check the stability of such a raft for a given set of parameters. In the first method, we use the boundary value solver FEniCS, which uses Newton's iteration to arrive at a solution. When we start with $\theta(r, 0) = 0$ as the initial guess, the solver converges to the positive solution $\theta_+(r)$. Then, we start the solver from an alternative initial condition, with $\theta_i(r, 0) = -\theta_+(r)$. In cases where there is a metastable solution corresponding to left-handed twist $\theta_-(r)$, the solver converges to this solution, while the solver instead converges to $\theta_+(r)$ if the counter-twisted raft is unstable.

As an independent check on the metastability of counter-twisted rafts, we also use the full $\hat{n}$ equations. We first generate an initial condition corresponding to a counter-twisted raft, by solving for the equilibrium configuration corresponding to preferred left-handed twist (i.e., we set $q_s = -q_s$). We then evolve this initial condition in time for preferred right-handed twist ($q_s > 0$) using the time-dependent Ginzburg Landau equations (described next). The stability of a counter-twisted raft is then determined by whether the dynamics reaches a steady-state corresponding to a right-handed or left-handed raft.

We performed these calculations as follows. Since we are only interested in the dynamics of how the rafts change twist, we freeze the concentration profile and time-evolve only the director field. We model the dynamics using the time-dependent Ginzburg Landau equations. Since the director field is not a conserved quantity, we use Model A dynamics (37), under the constraint that $\hat{n}(r)$ is a unit vector. This gives $\partial_t \hat{n} = -(\mathbf{I} - \hat{n}\hat{n}) \cdot \delta F / \delta \hat{n}$, with $\mathbf{I}$ the identity matrix. We calculate the director field dynamics on a uniformly discretized domain, using central difference approximations for the gradients and a classic Forward Euler implementation for the time-stepping (see the 'Finite Difference Numerical Method' section for details).

In addition to agreeing with the equilibrium calculations, the dynamics method corroborates the requirement for $K_1 > K_2$. In the regime where the raft does not untwist, if $K_1 = K_2$ the dynamics sometimes shows a different mechanism for the raft to flip back without untwisting, where the rods splay out and twist back to the preferred configuration. This instability is absent for $K_1 > K_2$. For this reason and the physical considerations described in the main text, we used $K_1 = 3K_2$ for all results described in this article. We will discuss the mechanisms of instability for counter-twisted rafts in detail in a future paper.

**Finite Difference Numerical Method:** To compute the dynamics of the director field, we consider a periodic domain $\Omega := [0, L] \times [0, L]$. The domain is discretized with a uniform mesh of spacing $h$ such that $L = Nh$. Thus, the points on the uniform mesh are given by $\mathbf{x}_{ij} = (x_i, y_j)^T$, where $x_i = 0 + ih$ and $y_j = 0 + jh$, and $1 \leq i, j \leq N$ are integers. We denote the scalar valued functions on this mesh as $f_{ij} = f(x_i, y_j)$. Now, the gradients are evaluated on this mesh using a finite central difference approximation:

$$\nabla_h f_{ij} := \left(\frac{1}{2h}(f_{i+1,j} - f_{i-1,j}), \frac{1}{2h}(f_{i,j+1} - f_{i,j-1})\right)^T \qquad [S5]$$

Other gradient operators including the curl and higher order terms are computed in a self-consistent manner. Finally, we use a uniform discretization of time $t_k := 0 + ks$, where $s > 0$ is the time step and $k$ is a positive integer. We now represent the fully discrete scalar function at $(x_i, y_j)$ at time $t_k$ by $f(t_k, x_i, y_j) := f_{ij}^k$ to simplify the notation.

The $\hat{n}$ dynamics is given by $\partial_t \hat{n} = -(I - \hat{n}\hat{n}) \cdot \delta F / \delta \hat{n}$. The Forward Euler scheme for this becomes:

$$\frac{\hat{n}_{ij}^{k+1} - \hat{n}_{ij}^k}{s} = -(I - \hat{n}_{ij}^k \hat{n}_{ij}^k) \cdot \left(\frac{\delta F}{\delta \hat{n}}\right)_{ij}^k \qquad [S6]$$

We used $L = 20$, in units of the twist penetration length, $N = 512$, and so $h \sim 0.04$. Since the highest spatial derivative in the equation is of second order, we used a time step $s \sim O(h^2)$, for the stability of the Forward Euler time stepping. For the single raft calculations, we placed the raft at the center of this domain using a tanh profile as described above. Denoting the mid-point by $\mathbf{x}_0 = \left(\frac{L}{2}, \frac{L}{2}\right)^T$, we have:

$$\boldsymbol{\phi}_{ij} = -\tanh\left(\frac{|\mathbf{x}_{ij} - \mathbf{x}_0| - R}{\sqrt{2\epsilon}}\right) \qquad [S7]$$

We used $R = 1$ and $\epsilon = 0.01$ (note that our grid spacing $h \sim 0.04$ is sufficiently small compared to the width of the tanh $\sqrt{2\epsilon} \sim 0.14$ and the radius of the raft). This $\phi$ profile puts a raft of short rods ($\phi = 1$) at the center of the box with a radius $R$, in a large background of long rods ($\phi = -1$). We use periodic boundary conditions for the box, and the large value of $L$ ensures that the raft does not interact with itself.

**Finite Element Calculation:** The differential equation to be solved for $\theta(r)$ is $\frac{\delta \beta F}{\delta \theta} = 0$, and hence:

$$-2\pi K_2 \frac{\mathrm{d}}{\mathrm{d}r}\left(r\frac{\mathrm{d}\theta}{\mathrm{d}r}\right) + \pi K_2 \sin 2\theta \left(r + \frac{1}{r} - r\,\gamma(\nabla\phi)^2\right) + 2\pi K_2\left(q + r\frac{\mathrm{d}q}{\mathrm{d}r}\right) - 2\pi K_2 q \cos 2\theta +$$
$$2\pi r\, C_2 \sin^2\theta \sin 2\theta = 0 \qquad [\text{S8}]$$

We note that $\frac{\mathrm{d}}{\mathrm{d}r}\left(r\frac{\mathrm{d}\theta}{\mathrm{d}r}\right) = r\nabla^2\theta$, and multiply the equation throughout by $-\frac{r}{2\pi}$ to get:

$$K_2 r^2 \nabla^2 \theta - \frac{1}{2} K_2 \sin 2\theta \left(r^2 + 1 - r^2\,\gamma(\nabla\phi)^2\right) - 2K_2 q \sin^2\theta - (r.\nabla q)r -$$
$$r^2\, C_2 \sin^2\theta \sin 2\theta = 0 \qquad [\text{S9}]$$

We have used $(1 - \cos 2\theta) = 2\sin^2\theta$ for simplifying the third term. Note that gradients in $\theta$ exist only in the first term. If we divide throughout by $K_2$ and denote the rest of the terms by $f(\theta)$, then the equation becomes:

$$r^2 \nabla^2 \theta + f(\theta) = 0 \qquad [\text{S10}]$$

The weak form of this equation for the finite element method is then given by:

$$\left(\nabla\theta, \nabla(r^2 v)\right) - (f(\theta), v) = 0 \qquad [\text{S11}]$$

We solved this equation in FEniCS (2, 3) in a circular domain of diameter $20\lambda$, with $\phi(r)$ set to the same tanh profile as before, with the raft at the center of this domain. We used an adaptive spatially varying mesh, with spacing h=0.2 far from the interface and finer mesh near the interface.

We imposed Dirichlet boundary conditions at the outer boundary of the domain. We used the non-linear solver in FEniCS, which derives the Jacobian matrix symbolically, and runs a Newton method to compute the solution. Due to the quartic nature of the equations, there is an unstable extremum close to $\theta \approx 0$ along with the two stable minima. To avoid the Newton iterator converging on the unstable solution, we used a non-zero positive twist (negative in case of counter-twisted rafts) as an initial guess. We used the `assemble()` function in FEniCS to calculate energies of the equilibrium configurations.

**Supplementary Figures**

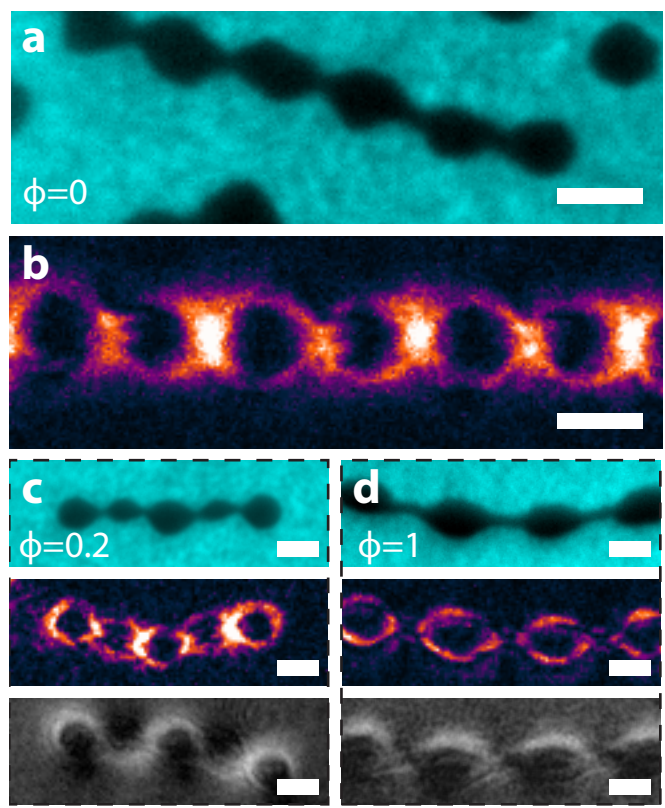

**Fig. S1: (a)** Short rods assemble into a chain of micron-sized rafts joined by narrow links of short rods in an achiral background. Here the long left-handed rods are fluorescently labeled. **(b)** PolScope images of chains in the achiral background show that twist is maximized at the inter-raft linkers. **(c)** In a low-chirality background, chains are formed of linked left- and right-handed rafts (imaged using tilted PolScope) with twist maximized between the links (imaged using PolScope). The left-handed links are smaller than right-handed links. **(d)** In a purely left-handed background, all rafts in the chain are the same size (imaged using fluorescence) and are right-handed (imaged using tilted PolScope) leading to longer necks where the twist decays to zero (PolScope).

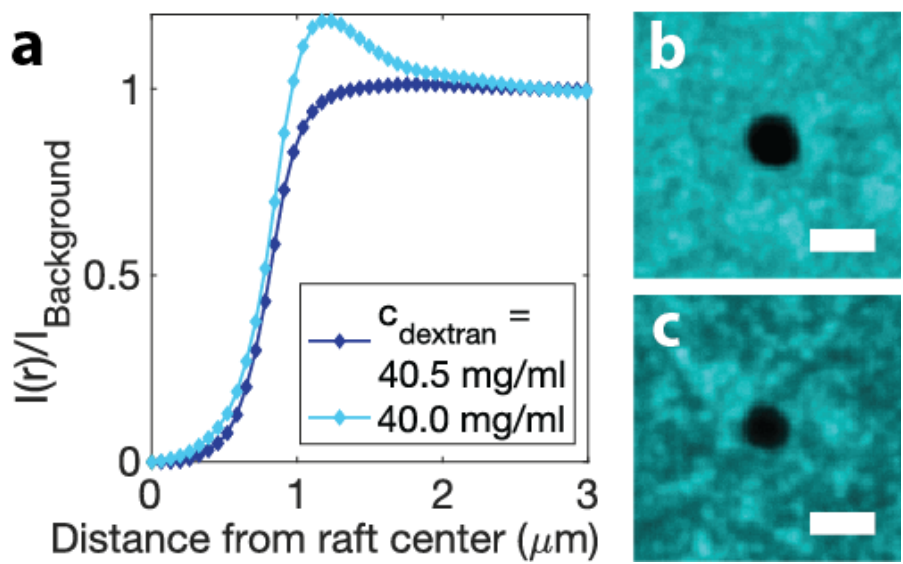

**Fig. S2: Small increases in depletant concentration lead to raft wetting. (a)** Radially- and temporally-averaged fluorescence intensity from labeled left-handed rods around rafts shows that at 40.5 mg/ml dextran concentration, the intensity of left-handed rods is higher at raft edges than in the raft or in the background, while at 40.0 mg/ml the intensity monotonically increases with distance from the raft. **(b)** At lower dextran concentrations, there is no noticeable wetting of left-handed rods at the raft edges, but **(c)** as dextran increases by less than 1% left-handed rods begin to wet the raft surface.

# R-R Pair Potential

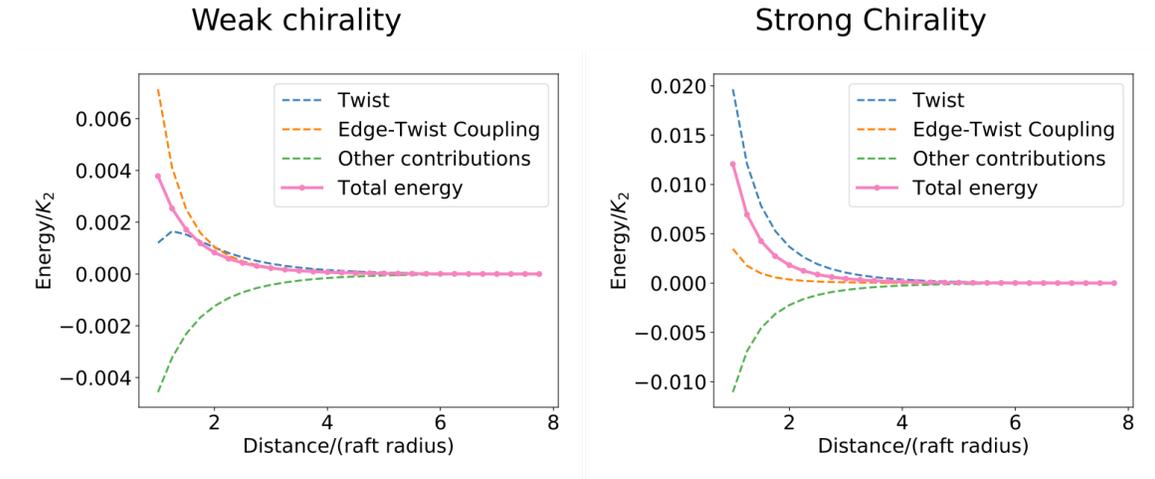

**Fig. S3:** Twist-dependent interfacial forces (labeled as 'Edge-Twist Coupling' in the legend) are dominant in the achiral limit, whereas the chiral forces are dominant for R-R raft repulsion in the chiral limit. The two plots show numerically calculated pair potentials for two right handed rafts. The parameters used are $K_1 = 3.0$, $\gamma = 0.8$ and $C_2 = 2000$, and (left panel) $q_r = 0.02$ and $q_b = 0.0$, and (right panel) $q_r = 0.6$ and $q_b = -0.4$.

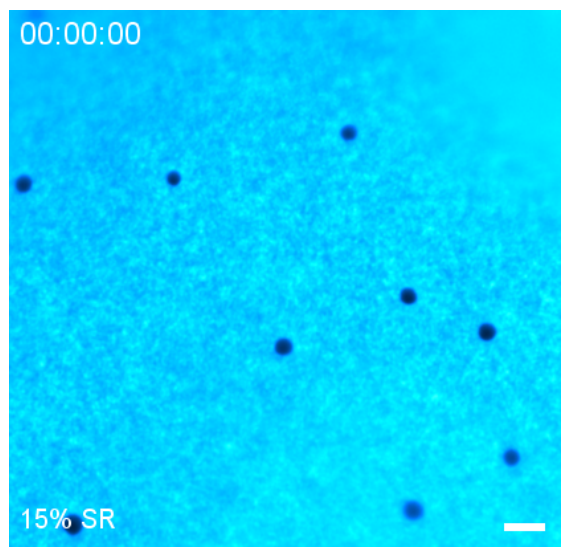

**Movie S1:** Time lapse imaging illustrates raft behavior in a membrane for three different stoichiometric ratios of long and short rods At low number density (15% short rods total) dilute rafts exhibit a gas-like structure. Rafts form higher-order structures such as tetramers and trimers in the membrane at intermediate raft density (25% short rods). At high density (30% short rods), rafts form square lattice-like assemblies leaving large open spaces in the membrane. Samples prepared at 40 mg/ml dextran concentration and the background membrane has no net chirality. The long left-handed rods are fluorescently labeled with DyLight488. Images were acquired using fluorescence and differential interference contrast (DIC) microscopy. Scale bar, 4 μm.

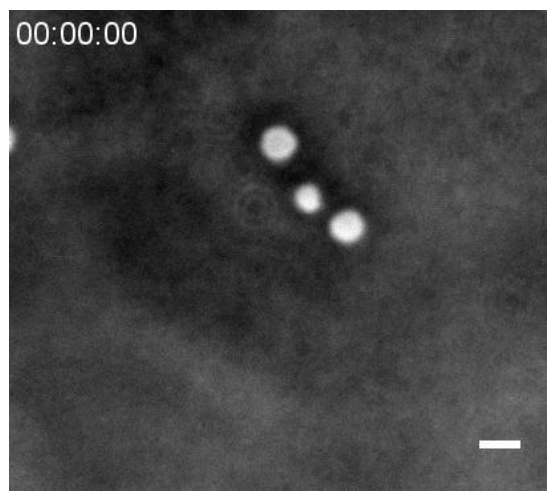

**Movie S2:** Phase contrast imaging time lapse shows short rod rafts assembled into a trimeric structure do not relax into an equilateral configuration in a colloidal membrane composed of an achiral mixture of left- and right-handed long rods. Scale bar is 2 μm.

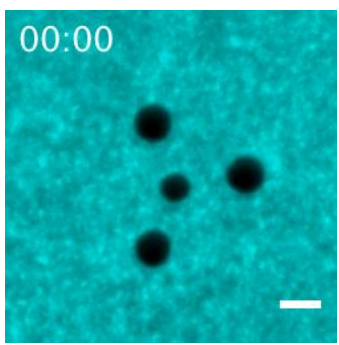

**Movie S3:** Colloidal rafts dissolved in an achiral membrane assemble into a stable tetrameric structure. Long left-handed rods are fluorescently labeled. Scale bar is 2 μm.

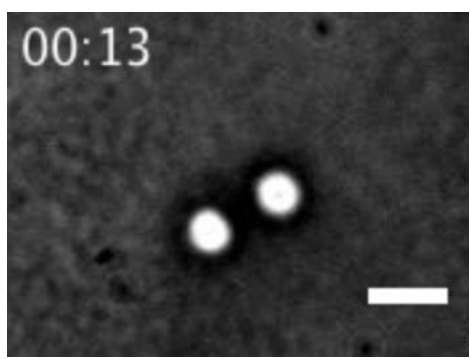

**Movie S4:** Phase contrast imaging of blinking optical trap technique demonstrates contrasting behavior of raft pairs in an achiral background membrane. Two optical plows made up of multiple time-shared light beams force two rafts together. The optical plows are switched off and rafts move away from each other. The first experimental run demonstrates repulsive raft interactions. The second experimental run shows colloidal rafts remaining bound together after the laser is shuttered, indicating attractive interactions. Scale bar, 2 μm.

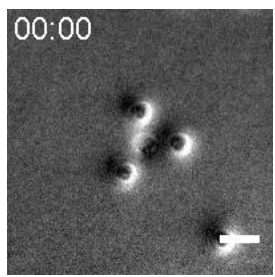

**Movie S5:** Tilted LC-PolScope time lapse of a tetrameric raft structure in an achiral background membrane demonstrates that the central raft has opposite twist from the outer rafts. Left-handed rafts exhibit a bright upper edge and dark lower edge, while right-handed rafts show a dark upper edge and bright lower edge. Scale bar, 2 μm.

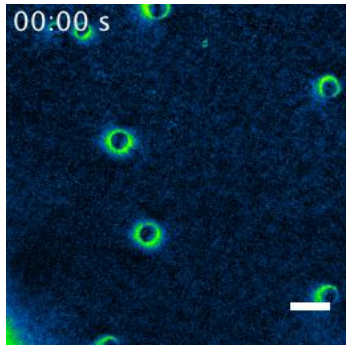

**Movie S6:** Normal and tilted LC-PolScope time lapse of left-handed rafts in an achiral background membrane of long rods shows the metastability of left-handed rafts. Normal incidence LC-PolScope imaging indicates that all rafts have similar tilt magnitude at their edges which remains stable over time. Tilted LC-PolScope shows that the two rafts on the center-left side are left-handed and thus counter-twisted. Scale bar, 2 µm.

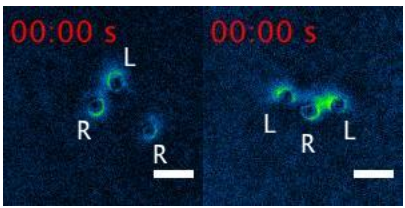

**Movie S7:** LC-PolScope and phase contrast imaging of two trimers assembled via optical plow in an achiral background membrane. Left: Trimer assembled from two right-handed rafts bound to a central left-handed raft. Right: Trimer assembled from two left-handed rafts around a central right-handed raft. Scale bar, 4 µm.

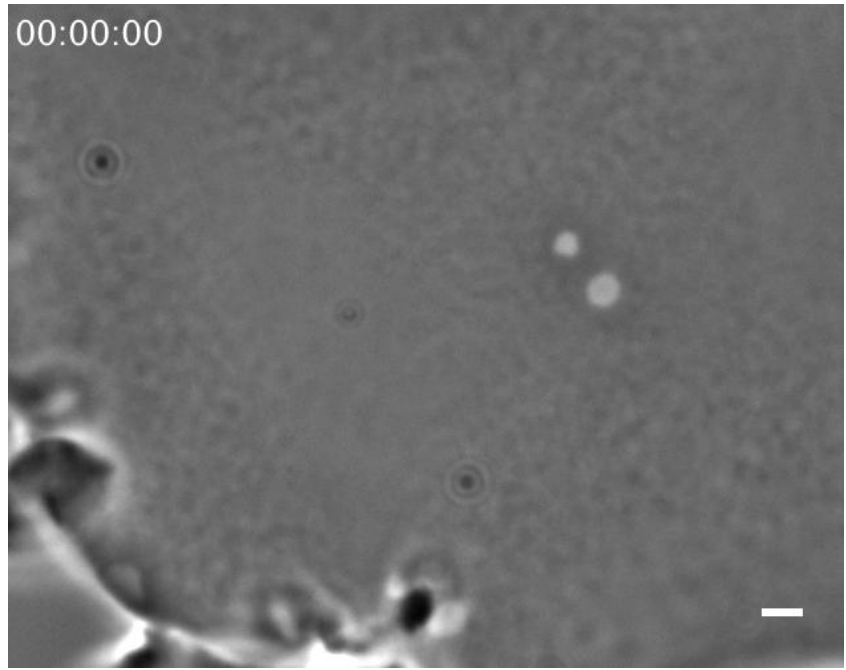

**Movie S8:** Time lapse imaging of a raft pair in a chiral background, $\phi_{ch} = 0.4$. The counter-twisted left-handed raft in the bound pair shrinks over a period of hours until the pair falls apart. At this point the shrunken raft begins to recover in size, indicating that it has switched to more favorable right-handed conformation. Scale bar, 2 μm.

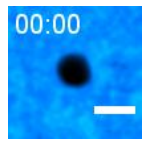

**Movie S9:** Long left-handed rods wet the raft edge at higher dextran concentrations. Fluorescence imaging shows that at 40 mg/ml Dextran concentration the long rod concentration is uniform constant around the raft edge, while at 40.5 mg/ml the labeled long, left-handed rods accumulate around the raft perimeter. Scale bar, 2 μm.